\documentclass[
   ,final            
  ]
  {aipproc}

\layoutstyle{6x9}

\begin{document}


\title{Search for New Phenomena in the CDF Top Quark Sample}


\author{Kevin Lannon}{
  address={The Ohio State University, Columbus, Ohio  43210, USA \\
           On Behalf of the CDF Collaboration}
}

\begin{abstract}
We present recent results from CDF in the search for new phenomena appearing in the top quark samples.  These results use data from $p\bar{p}$ collisions at $\sqrt{s} = 1.96$~TeV corresponding to an integrated luminosity ranging from 195~pb$^{-1}$ to 760~pb$^{-1}$.  No deviations are observed from the Standard Model expectations, so upper limits on the size of possible new phenomena are set.
\end{abstract}

\maketitle


\section{Introduction}
The top quark presents an intriguing opportunity in the search for new physics.  It is only recently that we have been able to study the top quark, and sample sizes remain relatively small. As a result, many properties of the top quark have not been measured well enough to verify they conform to Standard Model expectations.  The remarkable mass of the top quark (comparable to that of a gold nucleus) suggests that the top quark Yukawa coupling should be near unity, which may imply a special role for the top quark in electroweak symmetry breaking beyond the Standard Model.  In addition, there are many models for new physics which predict new phenomena in the top quark sector.  The results presented below use between 195~pb$^{-1}$ and 760~pb$^{-1}$ of $p\bar{p}$ data with $\sqrt{s} = 1.96$~TeV collected at the Tevatron Collider by the CDF experiment.

\section{Search for Top Decaying to a Charged Higgs}

\begin{figure}[htbp]
\includegraphics[width=.45\textwidth]{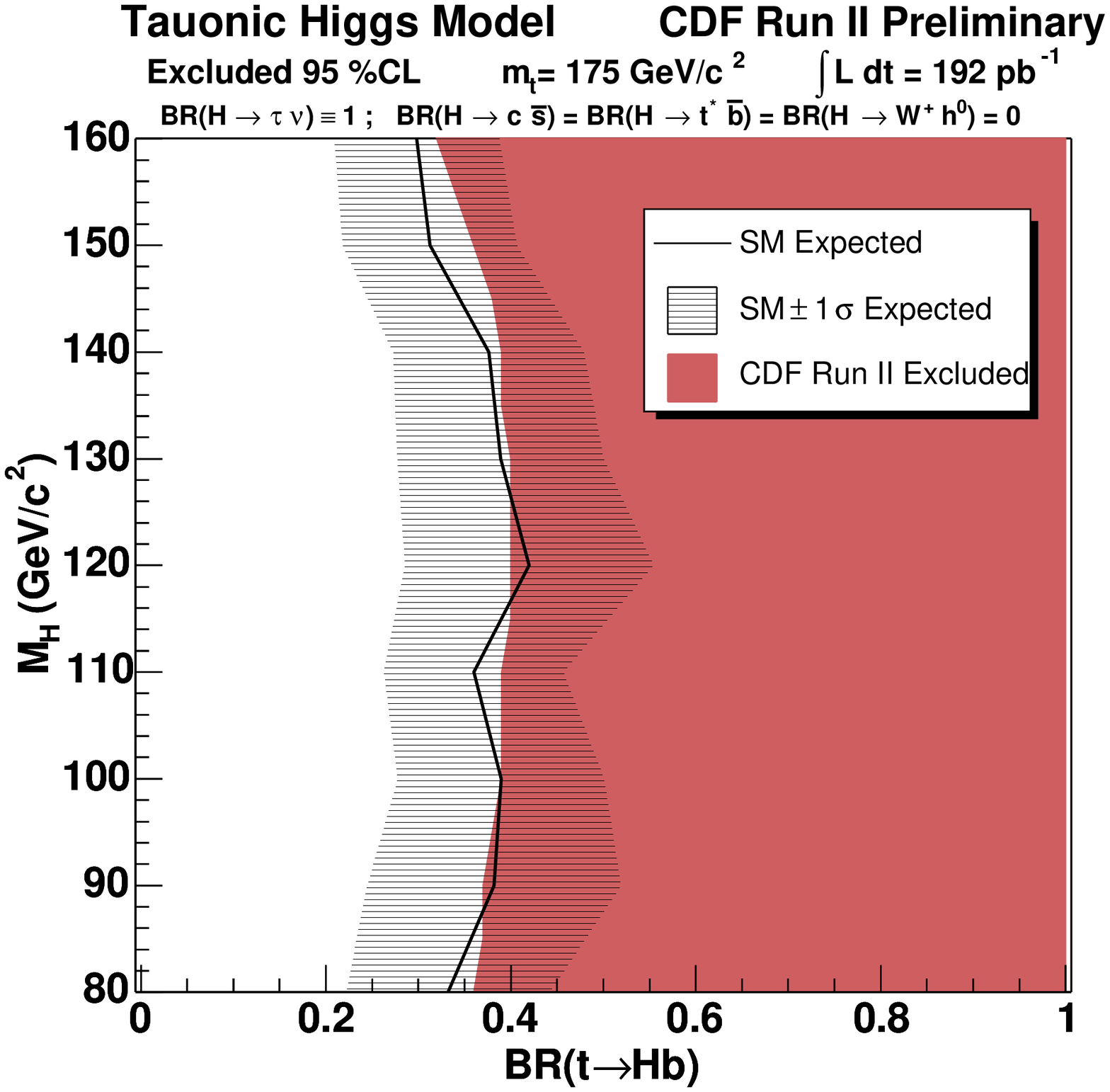}
\includegraphics[width=.45\textwidth]{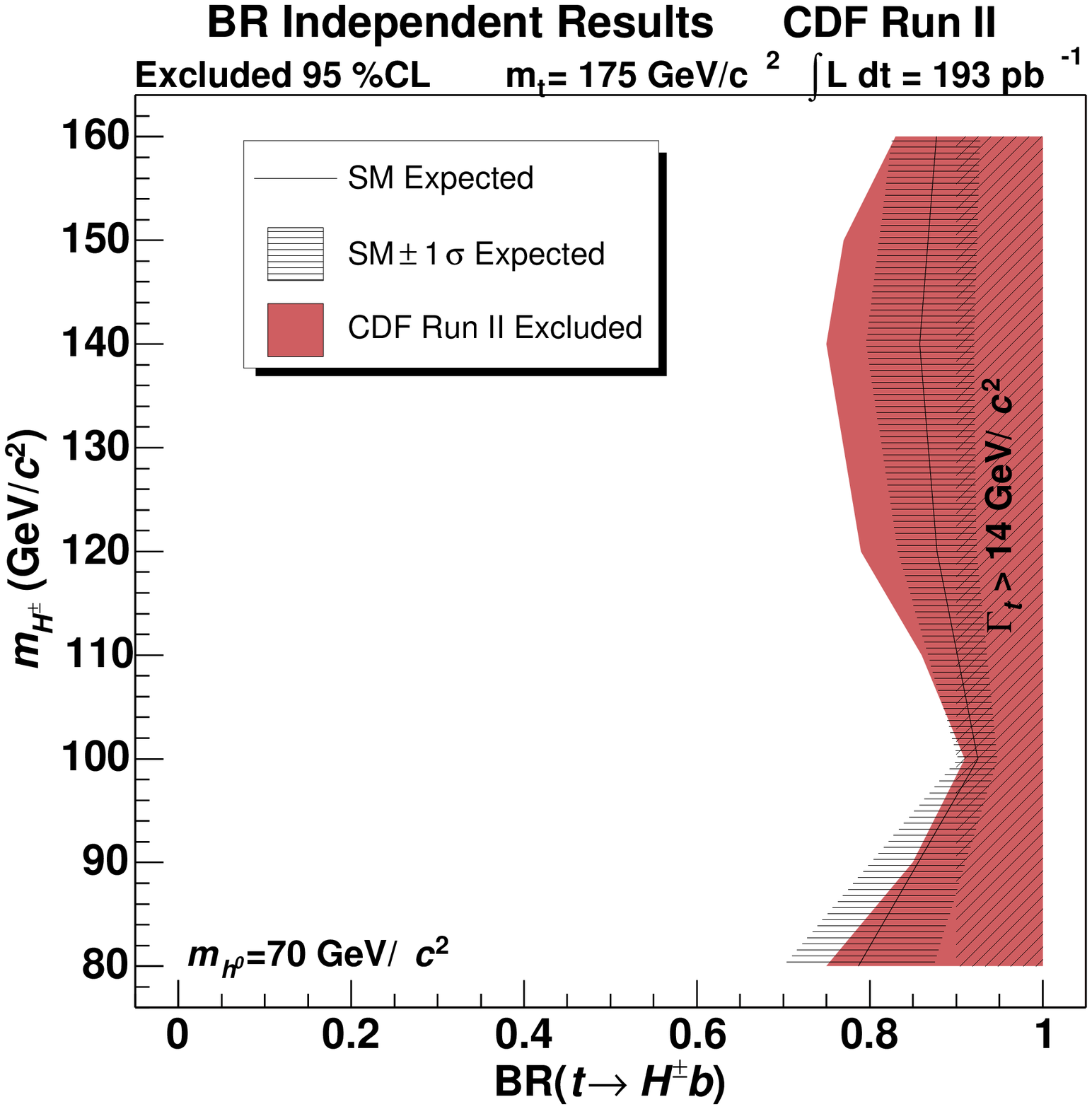}
\caption {\label{fig:chargedHiggs} The figure on the left show the excluded branching ratio of $t \to H^+b$ as a function of $M_{H^+}$ under the assumption that $BR(H^+ \to \tau\nu)= 1$, a common situation in MSSM models with high $\tan\beta$.  The figure on the right shows the branching ratio limit where, for each $M_{H^+}$, the $H^+$ branching ratios are chosen to give the least stringent limit.}
\end{figure}

Using 195 pb$^{-1}$ of data, we search for evidence of the decay $t \to H^+b$~\cite{Abulencia:2005jd} by comparing published CDF cross section measurements, made under the assumption that top decays only to $Wb$, in four different decay channels: dilepton~\cite{Acosta:2004uw}, lepton + jets with exactly one $b$-tag, lepton + jets with two or more $b$ tags~\cite{Acosta:2004hw}, and lepton + hadronically decaying $\tau$~\cite{Abulencia:2005et}.  Each of these results is individually consistent with Standard Model expectations.  However, allowing the top quark to decay via a charged Higgs can modify the expected yields in these final states, depending on the branching fractions of $t \to H^+b$ and allowed charged Higgs decay modes.  We calculate the branching ratio of $t \to H^+b$ and of $H^+ \to \tau\nu$, $cs$, $t^*b$, $W^+h^0$, and $W^+A^0$ for six different sets of minimal supersymmetric model (MSSM) parameters and use this information, combined with the observed yields in the four different decay modes, to set limits on the allowed values of $\tan\beta$ and $M_{H^+}$.  In addition, upper limits on $BR(t \to H^+b)$ as a function of $M_{H^+}$ are set under less model specific scenarios (see Fig.~\ref{fig:chargedHiggs}). 

\section{$W$ Helicity in Top Quark Decays}

In the Standard Model, the vector minus axial vector ($V-A$) structure of the electroweak interaction determines that approximately 70\% of $W$ bosons produced in top quark decays should have longitudinal polarization and the other 30\% should have left-handed polarization, assuming $M_{Top} = 175$ GeV/c$^2$.  Right-handed polarization is forbidden.  This $V-A$ structure can be tested by measuring the $W$ helicity angle, $\theta^*$, which is the angle between the negative direction of the top quark and the charged lepton from the $W$ decay measured in the $W$ rest frame.  Alternatively, the invariant mass of the charged lepton and the $b$ quark can be used, since $M_{lb} \approx 0.5(M_t^2-M_W^2)cos\theta^*$.  We employ both techniques to obtain the following results:  The fraction of logitudinal $W$ bosons, $F_0$ is measured, using a sample corresponding to an integrated luminosity of 320 pb$^{-1}$, to be $0.85^{+0.15}_{-0.22}\pm0.06$ assuming the fraction of right-handed $W$ bosons, $F_+$, is zero.  Using a 750 pb$^{-1}$ sample and assuming $F_0 = 0.7$, we establish an upper limit of $F_+ < 0.09$ at the 95\% confidence level.

\section{Top Quark Lifetime}

In the Standard Model, the top quark lifetime is on the order of 10$^{-24}$ seconds, which is too small to be measured by experiment.  However, there are several non-Standard-Model scenarios that could lead to a measurable lifetime in the top quark data sample.  For example, if top quarks were produced in the decay of some long-lived particle, or if some long-lived particle with a signature similar to the top quark had contaminated the sample.  With a 318 pb$^{-1}$ data sample, CDF measures the impact parameter distribution of leptons from the decay of $W$ bosons produced in top quark decays and sees no evidence of an anomalously long top quark lifetime, given the detector resolution and expected backgrounds.  We establish an upper limit on top quark lifetime of $c\tau < 52.5$~$\mu$m at the 95\% confidence level. 


\section {Search for a Resonance in the $t\bar{t}$ Invariant Mass Spectrum}

\begin{figure}[htbp]
\includegraphics[width=.45\textwidth]{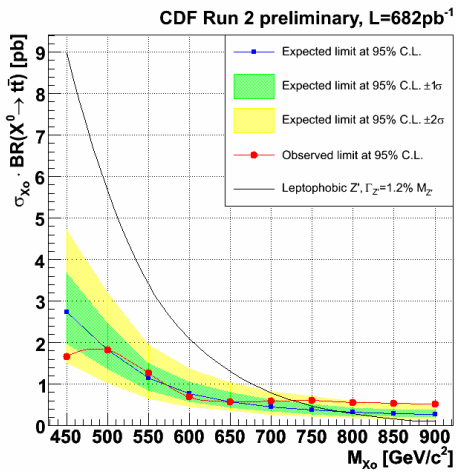}
\caption {\label{fig:mtt}  95\% CL upper limits for the cross section for resonance production. }
\end{figure}

Some models of physics beyond the Standard Model include new heavy particles that decay to top quark pairs.  For example, topcolor-assisted technicolor~\cite{Hill:1991at} predicts both topgluons and $Z^{\prime}$ that decay to $t\bar{t}$ pairs.  We preform a search for a generic, spin-1 resonance ($X_0$) with a narrow width ($\Gamma_{X_0}/M_{X_0} \approx 1.2\%$) decaying to top quark pairs by searching for deviations from the Standard Model predictions of the $t\bar{t}$ invariant mass spectrum.  No significant deviations are observed and we set 95\% confidence level upper limits on $\sigma_{X_0} \times BR(X_0 \to t\bar{t})$ (see Fig.~\ref{fig:mtt}).  The observed limits rule out a leptophobic $Z^\prime$ with mass less than 725 GeV/c$^2$.

\section {Search for $t^{\prime}$}

\begin{figure}[htbp]
\includegraphics[width=.5\textwidth]{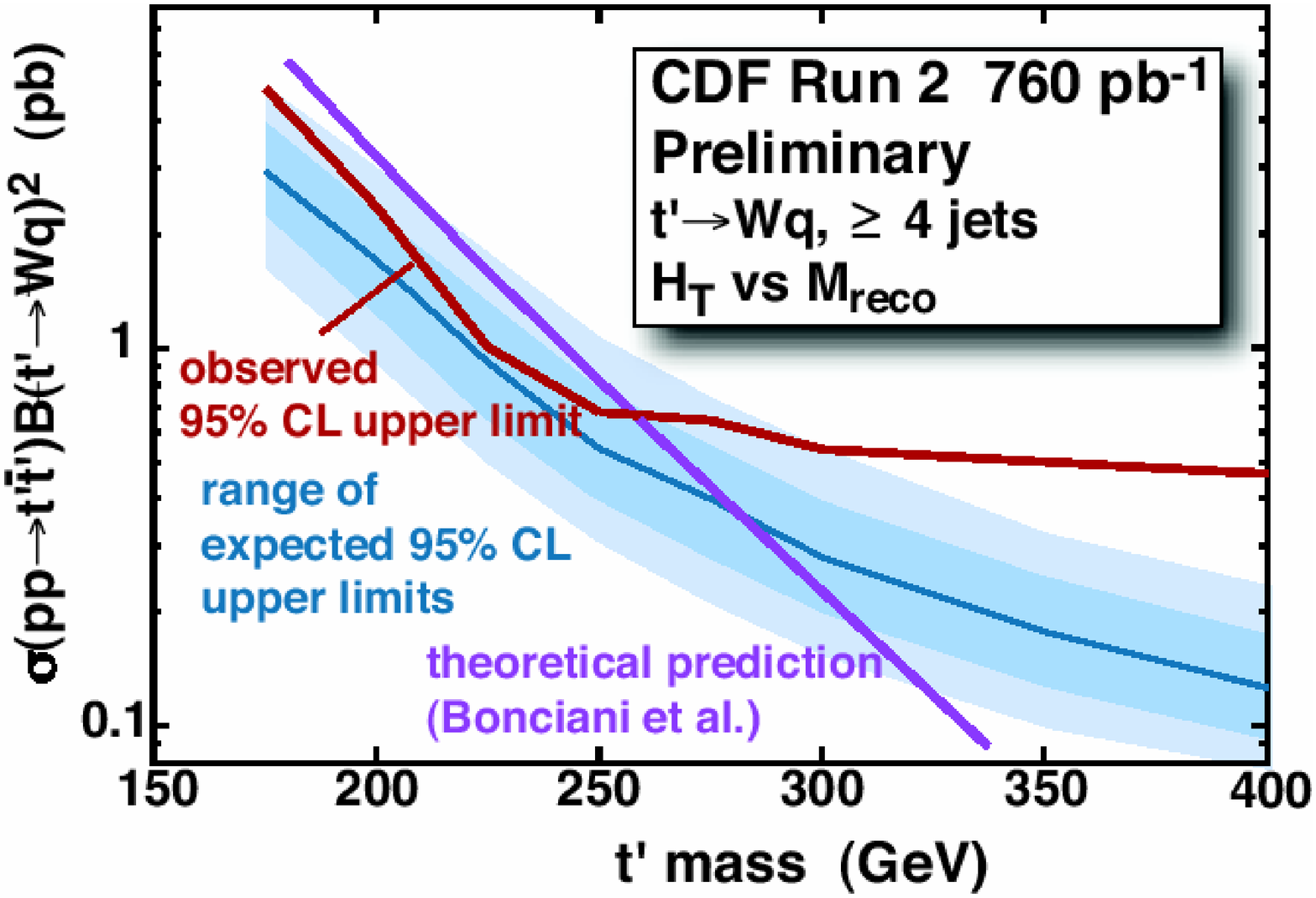}
\caption {\label{fig:tprime}  95\% CL upper limits for the cross section for $t^{\prime}$ production. }
\end{figure}

We search for evidence of a heavy, ``top-like'' object, $t^{\prime}$ in the CDF top quark sample.  Examples of models that predict such an object include a fourth chiral generation consistent with precision electroweak data~\cite{He:2001tp,Novikov:2001md} and the ``beautiful mirrors'' model~\cite{Choudhury:2001hs}, that predicts an additional quark generation that mixes with the third generation.  For this search, we assume that the $t^{\prime}$ has a large branching ratio to $Wq$, as would be the case if $M_{t^{\prime}} < M_{b^{\prime}} + M_W$, a situation favored by the constraint that an additional quark generation be consistent with precision electroweak data.  This search is performed using two kinematic variables to separate the $t^{\prime}$ signal from the Standard Model backgrounds: $H_T$, the sum of the transverse momenta of all objects in the event, and $M_{reco}$, the $Wq$ invariant mass reconstructed using the same $\chi^2$ fit technique employed by the CDF top mass analysis~\cite{Abulencia:2005aj}.  No evidence for $t^{\prime}$ is observed and we set 95\% confidence level upper limits on $\sigma_{t^{\prime}} \times BR(t^{\prime} \to Wq)^2$ that exclude $t^{\prime}$ masses below 258 GeV/c$^2$ (see Fig.~\ref{fig:tprime}).

\section{Conclusions}
Using up to 760 pb$^{-1}$ of data, we see no evidence for new phenomena in the CDF top quark sample.  However, ever larger data sets are rapidly becoming available.  There are several publications approaching completion using 1 fb$^{-1}$ or more of data, and several new analyses are nearing completion.  With a doubling time for the CDF sample of approximately one year, increasingly precise tests of the Standard Model using top quarks should be possible soon.





\bibliographystyle{aipproc}   

\bibliography{lannon_kevin}

\begin{thebibliography}{9}
\expandafter\ifx\csname natexlab\endcsname\relax\def\natexlab#1{#1}\fi
\providecommand{\enquote}[1]{``#1''}
\expandafter\ifx\csname url\endcsname\relax
  \def\url#1{\texttt{#1}}\fi
\expandafter\ifx\csname urlprefix\endcsname\relax\def\urlprefix{URL }\fi
\providecommand{\eprint}[2][]{\url{#2}}

\bibitem[Abulencia et~al.(2006{\natexlab{a}})]{Abulencia:2005jd}
A.~Abulencia, et~al., \emph{Phys. Rev. Lett.} \textbf{96}, 042003
  (2006{\natexlab{a}}), \eprint{hep-ex/0510065}.

\bibitem[Acosta et~al.(2004)]{Acosta:2004uw}
D.~Acosta, et~al., \emph{Phys. Rev. Lett.} \textbf{93}, 142001 (2004),
  \eprint{hep-ex/0404036}.

\bibitem[Acosta et~al.(2005)]{Acosta:2004hw}
D.~Acosta, et~al., \emph{Phys. Rev.} \textbf{D71}, 052003 (2005),
  \eprint{hep-ex/0410041}.

\bibitem[Abulencia et~al.(2006{\natexlab{b}})]{Abulencia:2005et}
A.~Abulencia, et~al., \emph{Phys. Lett.} \textbf{B639}, 172
  (2006{\natexlab{b}}), \eprint{hep-ex/0510063}.

\bibitem[Hill(1991)]{Hill:1991at}
C.~T. Hill, \emph{Phys. Lett.} \textbf{B266}, 419--424 (1991).

\bibitem[He et~al.(2001)]{He:2001tp}
H.-J. He, N.~Polonsky, and S.-f. Su, \emph{Phys. Rev.} \textbf{D64}, 053004
  (2001), \eprint{hep-ph/0102144}.

\bibitem[Novikov et~al.(2002)]{Novikov:2001md}
V.~A. Novikov, L.~B. Okun, A.~N. Rozanov, and M.~I. Vysotsky, \emph{Phys.
  Lett.} \textbf{B529}, 111--116 (2002), \eprint{hep-ph/0111028}.

\bibitem[Choudhury et~al.(2002)]{Choudhury:2001hs}
D.~Choudhury, T.~M.~P. Tait, and C.~E.~M. Wagner, \emph{Phys. Rev.}
  \textbf{D65}, 053002 (2002), \eprint{hep-ph/0109097}.

\bibitem[Abulencia et~al.(2006{\natexlab{c}})]{Abulencia:2005aj}
A.~Abulencia, et~al., \emph{Phys. Rev.} \textbf{D73}, 032003
  (2006{\natexlab{c}}), \eprint{hep-ex/0510048}.

\end{thebibliography}

\IfFileExists{\jobname.bbl}{}
 {\typeout{}
  \typeout{******************************************}
  \typeout{** Please run "bibtex \jobname" to optain}
  \typeout{** the bibliography and then re-run LaTeX}
  \typeout{** twice to fix the references!}
  \typeout{******************************************}
  \typeout{}
 }

\end{document}